%% file: 00_main.tex
\documentclass[aip,apl,amsmath,amssymb,reprint]{revtex4-1}

\draft 

\usepackage{graphicx}
\usepackage{dcolumn}
\usepackage{bm}

\usepackage[utf8]{inputenc}
\usepackage[T1]{fontenc}
\usepackage{mathptmx}
\usepackage{etoolbox}

\makeatletter
\def\@email#1#2{%
 \endgroup
 \patchcmd{\titleblock@produce}
  {\frontmatter@RRAPformat}
  {\frontmatter@RRAPformat{\produce@RRAP{*#1\href{mailto:#2}{#2}}}\frontmatter@RRAPformat}
  {}{}
}%
\makeatother

\usepackage{subfiles} 
\usepackage{orcidlink}

\usepackage{datetime}

\newcommand\tauD{\tau_\mathrm{d}}
\newcommand\tauR{\tau_\mathrm{r}}
\newcommand\tauPOne{\tau_\mathrm{p1}}
\newcommand\tauPTwo{\tau_\mathrm{p2}}

\DeclareSymbolFont{cmletters}{OMS}{cmsy}{m}{n}
\DeclareSymbolFontAlphabet{\mathcal}{cmletters}

\begin{document}

\subfile{10_title.tex}
\subfile{20_introduction.tex}
\subfile{30_methods.tex}
\subfile{40_experiment.tex}
\subfile{50_discussion.tex}
\subfile{60_before_appendix.tex}
\subfile{70_appendix.tex}

\nocite{*}
\bibliography{literature}

\end{document}

%% file: 10_title.tex
\title{Exponential-recovery model for free-running SPADs with capacity-induced dead-time imperfections}

\author{\orcidlinki{Jan~Krause}{0000-0002-3428-7025}}
\email{jan.krause@hhi.fraunhofer.de}
\affiliation{Fraunhofer Institute for Telecommunications, Heinrich-Hertz-Institut, HHI, 10587 Berlin, Germany}

\author{\orcidlinki{Nino~Walenta}{0000-0001-7243-0454}}
\affiliation{Fraunhofer Institute for Telecommunications, Heinrich-Hertz-Institut, HHI, 10587 Berlin, Germany}

\newdate{date}{14}{07}{2025}
\date{\displaydate{date}}

\begin{abstract}
Current count-rate models for single-photon avalanche diodes (SPADs) typically assume an instantaneous recovery of the quantum efficiency following dead-time, leading to a systematic overestimation of the effective detection efficiency for high photon flux.
To overcome this limitation, we introduce a generalized analytical count-rate model for free-running SPADs that models the non-instantaneous, exponential recovery of the quantum efficiency following dead-time.
Our model, framed within the theory of non-homogeneous Poisson processes, only requires one additional detector parameter -- the exponential-recovery time constant $\tauR$.
The model accurately predicts detection statistics deep into the saturation regime, outperforming the conventional step-function model by two orders of magnitude in terms of the impinging photon rate.
For extremely high photon flux, we further extend the model to capture paralyzation effects.
Beyond photon flux estimation, our model simplifies SPAD characterization by enabling the extraction of quantum efficiency $\eta_0$, dead-time $\tauD$, and recovery time constant $\tauR$ from a single inter-detection interval histogram.
This can be achieved with a simple setup, without the need for pulsed lasers or externally gated detectors.
We anticipate broad applicability of our model in quantum key distribution (QKD), time-correlated single-photon counting (TCSPC), LIDAR, and related areas.
Furthermore, the model is readily adaptable to other types of dead-time-limited detectors.
A Python implementation is provided as supplementary material for swift adoption.
\end{abstract}

\maketitle

%% file: 20_introduction.tex
\section{\label{sec:introduction}Introduction}

\begin{figure}[b]
    \includegraphics[width=\linewidth]{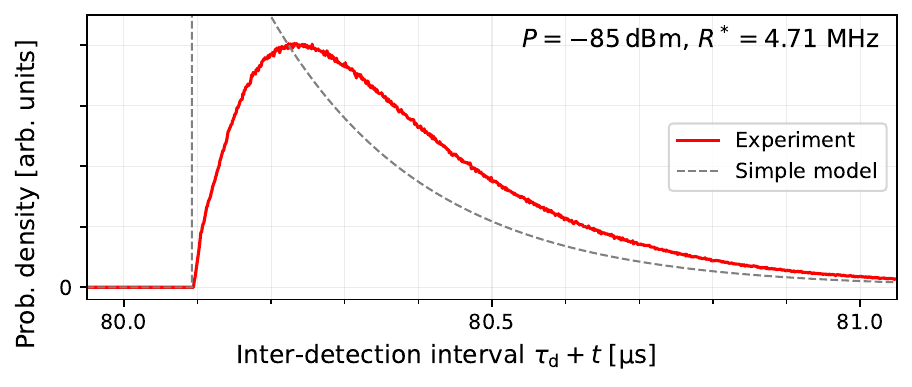}
    \caption{\label{fig:discrepancy}Discrepancy between simple model and experiment.
    The sharp rising edge at the end of the dead-time suggests that dead-time jitter is not prominent.
    The rounded shape around the maximum of the measured probability density suggests a non-instantaneous recovery of the SPAD quantum efficiency after the dead-time, likely caused by capacitive effects.
    }
\end{figure}

Single-photon avalanche diodes (SPADs) are the most widely used type of detector for single photon detection in the visible and near-infrared range
\cite{
    ghioniProgressSiliconSinglePhoton2007,
    itzlerAdvancesInGaAsPbasedAvalanche2011,
    ceccarelliRecentAdvancesFuture2021}.
They are used in a wide range of applications, including
quantum communication
\cite{
    gisinQuantumCryptography2002,
    zhangAdvancesInGaAsInP2015,
    scaraniSecurityPracticalQuantum2009},
quantum imaging
\cite{genoveseRealApplicationsQuantum2016},
and time-correlated single-photon counting (TCSPC)
\cite{beckerAdvancedTimeCorrelatedSingle2015},
just to name a few.

SPADs are avalanche photodiodes operated in the Geiger mode, i.e., with reverse bias above breakdown voltage.
An incoming photon excites a charge carrier with a probability $\eta_0$, called quantum efficiency, leading to a self-sustaining avalanche of charge carriers.
The resulting current is then typically amplified and processed electronically, e.g., by a time-to-digital converter (TDC).

After each detection event, the self-sustaining avalanche must be quenched to reset the SPAD and make it sensitive to incoming photons again.
This quenching is typically achieved via a passive quenching circuit
\cite{covaAvalanchePhotodiodesQuenching1996}.
Most free-running SPADs, i.e., those not gated externally, are operated with an additional latching circuit that actively keeps the reverse bias voltage below breakdown for a configurable time $\tauD$, called dead-time, after each detection event.
This is necessary to suppress false detections caused by afterpulsing, which results from the occupation of electron traps in the semiconductor material during the avalanche process
\cite{antiModelingAfterpulsingSinglephoton2011}.

Because the SPAD is not sensitive to photons during the dead-time window, the effective detection efficiency is reduced.
This effect is marginal for detection rates $R \ll \tauD^{-1}$, but becomes significant for higher detection rates.
The effect of reduced detection rate due to dead-time is described by the well-known relation (Ref.~\onlinecite{evansAtomicNucleus1955}, Ch. 28)
\begin{align}
\label{eq:simple_relation_1}
R
= \frac{1}{1/R^* + \tauD} \, ,
\end{align}
where $R$ is the measured detection rate, and $R^* = \eta_0 R_\mathrm{i}$ is the a priori detection rate, i.e., the virtual detection rate one would observe in the absence of dead-time.
$R_\mathrm{i}$ is the impinging photon rate and $\eta_0$ is the quantum efficiency.
The inverse relation is given by
\begin{align}
\label{eq:simple_relation_2}
R^*
= \frac{1}{1/R - \tauD} \, ,
\end{align}
and is often used to infer the impinging photon rate $R_\mathrm{i}$ or the quantum efficiency $\eta_0$ from the measured detection rate $R$.
Ultimately, the dead-time leads to a saturation effect, limiting the detection rate $R$ to $\tauD^{-1}$.

For the detector-on time $t$, i.e., the time between the end of the dead-time and the next detection, one expects for independent events a probability density function (PDF) of the form
\begin{align}
\label{eq:simple_pdf}
p(t) = R^* \, e^{-R^* t} \, .
\end{align}

\begin{figure}[t]
    \includegraphics[width=\linewidth]{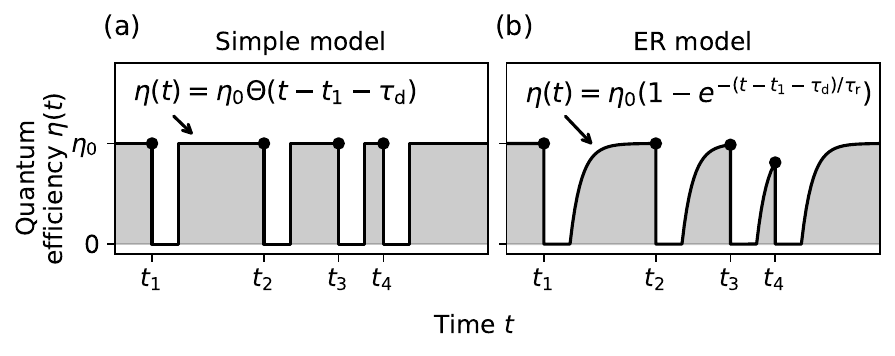}
    \caption{\label{fig:gates}Comparison of dead-time windows in the simple model and the exponential-recovery (ER) model.
    After a detection event at time $t_k$, the detector enters a dead-time of duration $\tauD$, during which it is insensitive to impinging photons.
    At time $t=t_k + \tauD$ the detector is activated again.
    Ideally, the dead-time windows have a perfect rectangular shape, shown in (a).
    For real devices, the quantum efficiency cannot rise back to $\eta_0$ instantaneously.
    We model the recovery by an exponential function, depicted in (b).
    This is motivated by the linear relation between quantum efficiency and excess bias voltage, which follows a capacitive charging behavior after each dead-time window.
    The recovery time constant is given by $\tauR = \mathcal{R} \, \mathcal{C}$, with typical quenching resistor values $\mathcal{R} \approx 100 \, \mathrm{k \Omega}$, and SPAD capacitance $\mathcal{C} \approx 1 \, \mathrm{pF}$, leading to $\tauR \approx 100 \, \mathrm{ns}$.
    }
\end{figure}

However, as shown in Fig.~\ref{fig:discrepancy}, this simple model does not agree with experimental data when the a priori detection rate $R^*$ is high.
Hence, fitting the simple model to experimental data leads to a significant underestimation of the impinging photon rate $R_\mathrm{i}$.
Furthermore, the rate equations \eqref{eq:simple_relation_1} and \eqref{eq:simple_relation_2} lose their validity.

The shape of the experimentally obtained distribution in Fig.~\ref{fig:discrepancy}, and the typical circuit design of SPADs
\cite{covaAvalanchePhotodiodesQuenching1996},
where the quenching resistor $\mathcal{R}$ and the diode capacity $\mathcal{C}$ lead to a capacitive charging behavior of the excess bias voltage after each dead-time window, suggest that the quantum efficiency does not recover instantaneously after the dead-time.

In this paper, we propose what we call exponential-recovery (ER) model, motivated by the aspects discussed above.
The ER model assumes a time-dependent quantum efficiency $\eta(t)$ after each dead-time window, described by an exponential recovery function.
This recovery function is completely characterized by the recovery time constant $\tauR$, an intrinsic detector property, independent of $R^*$.

In Section~\ref{sec:model_derivation}, we first derive a general approach for obtaining the PDF $p(t)$ for the detector-on time, as well as an adapted rate equation similar to Eq.~\eqref{eq:simple_relation_1} for any time-dependent detector efficiency $\eta(t)$.
Based on this approach, we then derive a closed-form PDF for the detector-on time, and adapted rate equations for an exponential recovery of the quantum efficiency after the dead-time.

Subsequently, we compare the model to experimental data in Section~\ref{sec:experiment}, finding excellent agreement for a wide range of a priori detection rates $R^*$, even for values as high as $R^* \approx 10 \, \tauR^{-1}$.
For higher values of $R^*$, additional effects, like detector paralyzation, dominate the measured rate $R$.
We capture this behavior by a paralyzing extension of our ER model in Section~\ref{sec:paralyzing_model}.
We conclude in Section~\ref{sec:discussion} by providing multiple ideas for applications and extensions of our model.

%% file: 30_methods.tex
\section{\label{sec:model_derivation}Theoretical model derivation}

\begin{figure}[t]
    \includegraphics[width=0.9\linewidth]{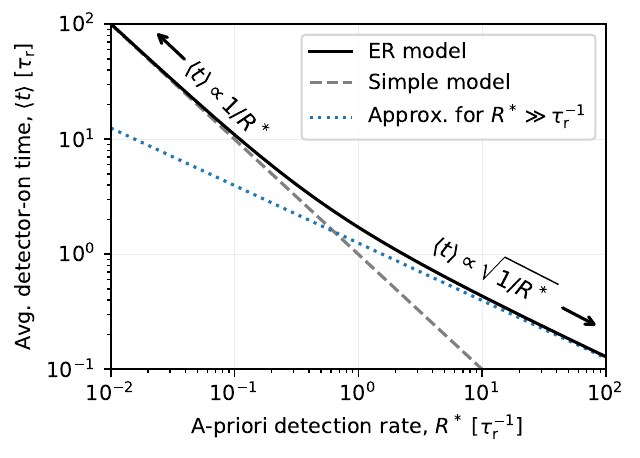}
    \caption{\label{fig:model_comparison}Model comparison.
    The predicted average detector-on time $\langle t \rangle$ as a function of the a priori detection rate $R^*$ is shown by the dashed gray (solid black) curve for the simple model (exponential-recovery (ER) model).
    The dotted blue curve is the lowest order approximation for $R^* \gg \tauR^{-1}$, derived in Appendix~\ref{app:exact_model_approximation_high_rates}.
    For increasing $R^*$ the power law of $\langle t \rangle$ changes from $\langle t \rangle = 1/R^*$ to $\langle t \rangle = \sqrt{\pi \tauR / 2 R^*}$, cf. Appendix~\ref{app:exact_model_approximation_high_rates}.
    }
\end{figure}

\begin{table}[b]
    \caption{\label{tab:nomenclature}Notation used throughout the text.}
    \begin{ruledtabular}
        \begin{tabular}{p{0.16\linewidth}p{0.82\linewidth}}
            Symbol & Meaning \\
            \hline
            $\eta_0$ & \textit{Asymptotic quantum efficiency}, approached for long detector-on times, cf. Fig.~\ref{fig:gates} \\
            $R_\mathrm{i}$ & \textit{Impinging photon rate}, i.e., the rate at which photons hit the detector \\
            $R^* = R_\mathrm{i} \, \eta_0$ & \textit{A priori detection rate}, i.e., the virtual detection rate one would observe in the absence of dead-time windows and for instantaneous recovery of the quantum efficiency \\
            $R$ & Experimentally observed \textit{detection rate} \\
            $\tauD$ & \textit{Dead-time} window duration \\
            $\tauR$ & \textit{Exponential-recovery time constant} of the quantum efficiency recovery after a dead-time window, cf. Fig.~\ref{fig:gates} \\
            $\tauPOne$ & Paralyzing ER model \textit{paralyzable-time constant} \\
            $\tauPTwo$ & Paralyzing ER model \textit{dead-time extension constant}
        \end{tabular}
    \end{ruledtabular}
\end{table}

In this section, we first introduce a general approach for deriving the PDF, $p(t)$, of the detector-on time for any time-dependent detector efficiency $\eta(t)$.
The PDFs for the simple model and the exponential-recovery (ER) model are then easily obtained as a special case by plugging the respective $\eta(t)$ into Eq.~\eqref{eq:general_pdf}.
The resulting PDF can then readily be used to fit experimental data and to check the validity of the assumed $\eta(t)$.
Finally, calculation of the average detector-on time, Eq.~\eqref{eq:t_mean_general}, leads to the general rate equation Eq.~\eqref{eq:general_rate_equation}.

An overview over the most relevant variables is given in Table~\ref{tab:nomenclature}.

\subsection{General framework for non-instantaneous recovery}
\label{sec:general_framework}

Throughout this work, it is assumed that the SPAD is operated in the free-running mode, and is illuminated by a constant-power continuous-wave (CW) light source, such as a laser, for which the photon number distribution is Poissonian.
The detector is assumed to be non-paralyzing, i.e., the dead-time duration is assumed fix and independent of the photons hitting the detector.
Laser‐intensity noise and afterpulsing are not considered.
Furthermore, we use $t=0$ for the end of the dead-time window for simplicity.

The complementary cumulative distribution function (CCDF), $\bar{P}(t) = 1 - P(t)$, which is the probability to have no detection since the end of the last dead-time window until time $t$, can be written in discretized form as
\begin{align}
\bar{P}(t) = \bar{P}(0;t) = \prod_{k=1}^n \bar{P} \pmb{\big(} (k-1) \Delta t; k \Delta t \pmb{\big)} \, ,
\end{align}
where $\bar{P}(t_1; t_2)$ is the probability for no detection in the interval $[t_1, t_2]$ condition on having no detection in the preceding interval $[0, t_1]$, and $\Delta t = t/n$.
For small $\Delta t$ one finds
\begin{align}
\label{eq:short_interval_p_bar}
\bar{P} \big(t; t + \Delta t \big)
&\approx e^{-R_\mathrm{i} \eta(t) \Delta t} \, ,
\end{align}
where a Poissonian photon number distribution is assumed.
Combination of both equations leads to
\begin{align}
\bar{P}(t)
&\approx \exp \pmb{\Big(} -R_\mathrm{i} \sum_{k=1}^n \eta(k \Delta t) \Delta t \pmb{\Big)} \, .
\end{align}
In the limit $\Delta t \to 0$ the approximation in Eq.~\eqref{eq:short_interval_p_bar} becomes exact and one finds
\begin{align}
\label{eq:cdf}
P(t)
&= 1 - \bar{P}(t)
= 1 - \exp \pmb{\Big(} -R_\mathrm{i} \int_{0}^t \eta(\bar{t}) \, \mathrm{d}\bar{t} \pmb{\Big)} \, .
\end{align}
The PDF is now obtained as
\begin{equation}
\boxed{
p(t)
= \frac{\mathrm d}{\mathrm d t} P(t)
= R_\mathrm{i} \, \eta(t) \exp \pmb{\Big(} -R_\mathrm{i} \int_{0}^t \eta(\bar{t}) \, \mathrm{d}\bar{t} \pmb{\Big)} \, . \label{eq:general_pdf}
}
\end{equation}
Normalization of this PDF is implied by Eq.~\eqref{eq:cdf}, as long as the integral in the exponent diverges for $t \to \infty$.
For a specific time-dependent detector efficiency $\eta(t)$, this PDF can readily be used to fit experimental data.

Note, that our model describes in fact a non-homogeneous Poisson process (NHPP)
\cite{snyderRandomPointProcesses1991},
which has the general PDF
\begin{align}
p(t)
= \lambda(t) \exp \pmb{\Big(} - \int_{0}^t \lambda(\bar{t}) \, \mathrm{d}\bar{t} \pmb{\Big)} \, , \label{eq:pdf_non_homogeneous_poisson}
\end{align}
for a time-dependent event rate $\lambda(t)$.

To derive a rate equation similar to Eq.~\eqref{eq:simple_relation_1}, the average detector-on time
\begin{align}
\label{eq:t_mean_general}
\langle t \rangle
&= \int_0^\infty t p(t) \, \mathrm{d} t
\end{align}
can be plugged into the general rate equation
\begin{align}
\label{eq:general_rate_equation}
R = \frac{1}{\langle t \rangle + \tauD} \, .
\end{align}

\subsection{Simple model for instantaneous recovery}

For instantaneous recovery of the detector efficiency after the dead-time, the detector efficiency is given by
\begin{align}
\eta(t) = \eta_0 \, .
\end{align}
Together with Eq.~\eqref{eq:general_pdf} this directly leads to the PDF from Eq.~\eqref{eq:simple_pdf}, and the average detector-on time $\langle t \rangle = 1/R^*$, allowing for the recovery of Eq.~\eqref{eq:simple_relation_1}.

\subsection{Exponential-recovery (ER) model}

\begin{figure*}
    \includegraphics[width=\linewidth]{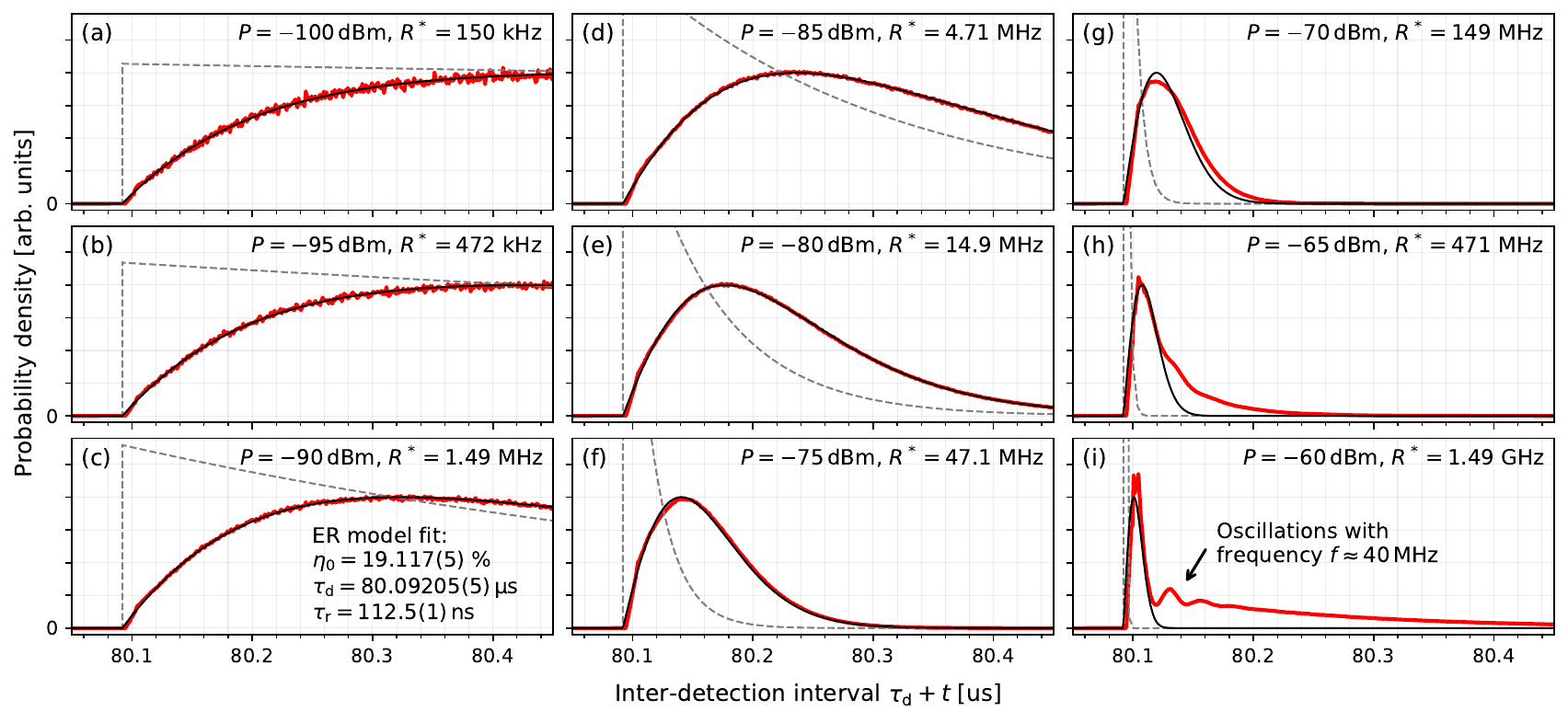}
    \caption{\label{fig:experiment_combined}Comparison of inter-detection interval distributions for experiment and ER model.
    Shown are the plots for optical power incident on the detector between $-100$ and $-60\,\mathrm{dBm}$ at a wavelength of $1546.92 \, \mathrm{nm}$.
    These powers correspond to impinging photon rates of $R_\mathrm{i} = 7.79 \times 10^{5}$ to $7.79 \times 10^{9} \, \mathrm{s}^{-1}$.
    The red curves show the data, binned into 1-ns bins.
    All three parameters, the asymptotic quantum efficiency $\eta_0$, the dead-time $\tauD$, and the exponential-recovery time constant $\tauR$ were fitted to the data for an impinging photon rate of $1.49 \, \mathrm{MHz}$, see black line in (c).
    For the other datasets, $\eta_0$, $\tauD$ and $\tauR$ were fixed to the previously fitted values, and only an additional vertical scaling factor was fitted to account for the onset of detector paralyzation for high impinging photon rates $R^*$ in (g)-(i).
    The dashed gray curves depict the simple model for the calibrated a priori detection rates $R^*$.
    The ER model is in very good agreement with the data for a priori detection rates as high as $R^* = 47.1 \, \mathrm{MHz}$.
    Above this value, additional effects come into play, which are not accounted for by the ER model, see Section~\ref{sec:limitations}.
    To simplify visual comparison, data and fit were vertically scaled individually for each subfigure, such that the fit maxima coincide for all plots.
    }
\end{figure*}

To obtain a more precise model for a time-dependent quantum efficiency after each dead-time window, we use the fact that the quantum efficiency is approximately proportional to the excess bias voltage
\cite{covaAvalanchePhotodiodesQuenching1996},
which exhibits a capacitive recharge behavior after each dead-time window
\cite{covaAvalanchePhotodiodesQuenching1996,
    sanzaroInGaAsInPSPAD2016,
    raupachDetectionRateDependence2022}.
This recharge behavior is determined by the quenching resistor resistance $\mathcal{R}$, and the diode capacity $\mathcal{C}$, leading to the recovery time constant $\tauR = \mathcal{R} \, \mathcal{C}$.
The time-dependent quantum efficiency is then given as
\begin{align}
\eta(t) = \eta_0 \, (1 - e^{-t / \tauR}) \, ,
\end{align}
where $\eta_0$ is the asymptotic quantum efficiency, i.e., the quantum efficiency approached for long detector-on times, cf. Fig.~\ref{fig:gates}.
Using Eq.~\eqref{eq:general_pdf}, this leads to the PDF
\begin{equation}
\label{eq:exact_model_pdf}
\boxed{
p(t)
= \Big[ e^{ R^* \tauR \, (1 - e^{-t/\tauR}) } \big( 1 - e^{-t/\tauR} \big) \Big] \times R^* e^{-R^* t}
}
\end{equation}
of the detector-on time, where the term in the square brackets can be seen as the correction to the simple model.
This PDF can readily be fitted to experimental data, cf. Fig.~\ref{fig:experiment_combined}.

The average detector-on time
\begin{align}
\label{eq:t_mean_er_model}
\langle t_\textrm{ER model} \rangle
&\equiv \langle t \rangle
= \int_0^\infty t p(t) \, \mathrm{d} t
\end{align}
does not possess a closed form solution.
Therefore, the adapted rate equation, Eq.~\eqref{eq:general_rate_equation}, is evaluated numerically for the remainder of this paper.
However, analytical approximations of $\langle t \rangle$ can be obtained and are provided for $R^* \ll \tauR^{-1}$ and $R^* \gg \tauR^{-1}$ in the Appendix~\ref{app:exact_model_approximation_low_rates} and \ref{app:exact_model_approximation_high_rates}, respectively.

A Python implementation of the numerical evaluation and the analytical approximations is provided as supplementary material.

\begin{figure}
    \includegraphics[width=\linewidth]{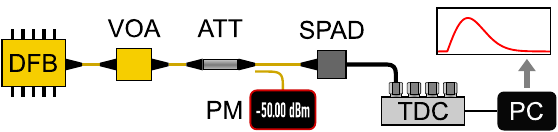}
    \caption{\label{fig:setup}Experimental setup.
    The simple setup works without pulsed lasers or synchronized detector gating.
    Fitting Eq.~\eqref{eq:exact_model_pdf} to a single inter-detection interval histogram allows for the extraction of the asymptotic quantum efficiency $\eta_0$, the dead-time $\tauD$, and the exponential-recovery time constant $\tauR$.
    ATT: Fixed attenuator;
    DFB: Distributed-feedback laser;
    PC: Personal computer;
    PM: Powermeter;
    SPAD: Single-photon avalanche diode;
    TDC: Time-to-digital converter;
    VOA: Variable optical attenuator.
    }
\end{figure}

\subsection{Model comparison}

The impact of the exponential recovery function becomes relevant when the a priori detection rate $R^*$ approaches $\tauR^{-1}$.
The effect on the average detector-on time $\langle t \rangle$ is visualized in Fig.~\ref{fig:model_comparison}.
For $R^* \ll \tauR^{-1}$, the average detector-on time is well approximated by the simple model, i.e., $\langle t \rangle \approx 1/R^*$, while for $R^* \gg \tauR^{-1}$ the relation changes its power law, approaching $\langle t \rangle \approx \sqrt{\pi \tauR / 2 R^*}$.

%% file: 40_experiment.tex
\section{\label{sec:experiment}Experiment}

We tested our model on experimental data, finding excellent agreement for a wide range of a priori detection rates $R^*$, even up to $R^* \approx 10 \, \tauR^{-1}$, see Fig.~\ref{fig:experiment_combined}.

To obtain datasets for different values of $R^*$, we illuminated a commercial SPAD (ID Quantique, IDQube-NIR-FR-MMF-LN) using a distributed-feedback (DFB) laser at a wavelength of $1546.92 \, \mathrm{nm}$, attenuated by a variable optical attenuator (VOA) and a $60\,\mathrm{dB}$ fixed attenuator, see Fig.~\ref{fig:setup}.
For $0\,\mathrm{dB}$ VOA attenuation, the optical power incident on the SPAD was measured using a calibrated powermeter (S154C, Thorlabs) with a relative measurement uncertainty of 5\%.
For higher VOA attenuations, the linearity of the VOA was measured and found to be within $0.5\,\mathrm{dB}$ over its whole range.
The SPAD temperature was set to $-40^\circ \mathrm{C}$, the efficiency to $20\,\%$, and the dead-time to the maximum value of $\tauD = 80\,\mathrm{\mu s}$ to minimize the impact of afterpulsing.

Timestamps were recorded by a time-to-digital converter (Swabian Instruments, TimeTagger Ultra) with a standard deviation of $8\,\mathrm{ps}$.
For data analysis, the inter-detection intervals were binned into $1\,\mathrm{ns}$ bins.
To keep the statistical uncertainty low, at least $10^7$ timestamps were recorded for each VOA setting.
Dark-counts were taken into account in all cases where detection rates were processed, by adding the a priori dark-count rate of $858\,\mathrm{Hz}$, obtained via Eq.~\eqref{eq:simple_relation_2}, to the a priori count-rates.

We recorded datasets for powers from $-110$ to $-50 \, \mathrm{dBm}$ incident on the detector.
This power range corresponds to impinging photon rates of $R_\mathrm{i} = 7.79 \times 10^{4} \,\, \text{--} \,\, 7.79 \times 10^{10} \, \mathrm{s}^{-1}$.

The resulting histograms and fits of the ER model for incident powers between $-100$ and $-60\,\mathrm{dBm}$ are shown in Fig.~\ref{fig:experiment_combined}.
The three model parameters $\eta_0$, $\tauD$, and $\tauR$ were fitted to the dataset in Fig.~\ref{fig:experiment_combined}(c), leading to $\eta_0 = 19.117(5)\,\%$, $\tauD = 80.09205(5) \, \mathrm{\mu s}$ and $\tauR = 112.5(1) \, \mathrm{ns}$.
For all other datasets in Fig.~\ref{fig:experiment_combined}, only a vertical scaling factor was fitted, while fixing $\eta_0$, $\tauD$, and $\tauR$ to the previously fitted values.
The model shows excellent agreement with the data for a priori detection rates as high as $R^* \approx 50 \, \mathrm{MHz}$, cf. Fig.~\ref{fig:experiment_combined}(f).
For higher values of $R^*$, the model still provides a good fit to the data, but deviates in the right tail of the PDF, as shown in Fig.~\ref{fig:experiment_combined}(g)-(i).
These model limitations are further discussed in Section~\ref{sec:limitations} and addressed by a paralyzing extension of the ER model in Section~\ref{sec:paralyzing_model}.

\subsection{Inferred a priori detection rates}
\label{sec:inferred_rates}

\begin{figure}[t]
    \includegraphics[width=\linewidth]{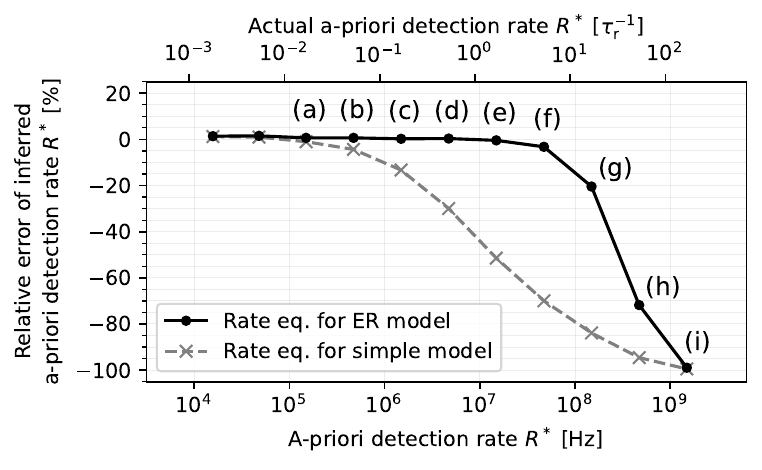}
    \caption{\label{fig:inferred_rates}Comparison of simple model and ER model in terms of the accuracy of inferred a priori detection rates.
    The dashed gray line depicts the results obtained from the rate equation of the simple model, Eq.~\eqref{eq:simple_relation_2}.
    The solid black line is obtained, when using the general rate equation, Eq.~\eqref{eq:general_rate_equation}, where the average detector-on time $\langle t \rangle$, Eq.~\eqref{eq:t_mean_er_model}, was numerically evaluated for the exponential-recovery (ER) model without approximations.
    The labels refer to Fig.~\ref{fig:experiment_combined}.
    }
\end{figure}

It is of practical interest to infer the a priori detection rate $R^*$ from the measured detection rate $R$, cf. Ref.~\onlinecite{lopezStudyDevelopRobust2020}.
Therefore, the results for the adapted rate equation, Eq.~\eqref{eq:general_rate_equation}, are compared with the rate equation of the simple model in Fig.~\ref{fig:inferred_rates}.
The ER model is in excellent agreement with the data for a priori detection rates as high as $R^* = 47 \, \mathrm{MHz}$, where we find a relative deviation of only $-3.23\,\%$ between the inferred and the actual $R^*$.
The simple model achieves comparable precision only up to $R^* = 0.47 \,\mathrm{MHz}$.
Hence, our ER model extends the applicable a priori detection range by a factor of approx. 100.

\subsection{Fit stability}
\label{sec:fit_stability}

To verify the assumption that $\tauR$ is an intrinsic detector property, we fitted $\tauR$ and a vertical scaling parameter to the data for each a priori detection rate $R^*$, see Fig.~\ref{fig:fit_stability}.
The values for $\eta_0$ and $\tauD$ were fixed to the values obtained for the dataset in Fig.~\ref{fig:experiment_combined}(c).
The fitted values of $\tauR$ deviate by less than $4.36\,\%$ from point (c) for a priori detection rates up to $R^* = 47.1 \, \mathrm{MHz}$, confirming the assumption that $\tauR$ is an intrinsic detector property.
A comparable fit stability is also observed, when keeping $\eta_0$ and $\tauD$ as additional free fit parameters for every dataset.
The deviations for higher values of $R^*$ most likely stem from the effects described in the following.

\subsection{Limitations of the ER model}
\label{sec:limitations}

For a priori detection rates above $R^* = 47.1 \, \mathrm{MHz}$, the ER model begins to deviate from the data.
As can be seen in Fig.~\ref{fig:experiment_combined}(g)-(i), three new effects emerge for high values of $R^*$:
First, the data for the right tail of the PDF lies clearly above the PDF of the ER model.
Second, the right tail of the PDF exhibits oscillations with a frequency $f \approx 40\,\mathrm{MHz}$, see Fig.~\ref{fig:experiment_combined}(i).
Third, on the left side of the PDF, the histogram values consistently lie below the model prediction for a duration of approx. $10\,\mathrm{ns}$, independent of $R^*$.
All three effects are not accounted for by the ER model.
Furthermore, the detection rate $R$ begins to decrease for increasing $R^*$, suggesting the onset of detector blinding, i.e., the transition from a non-paralyzing to a paralyzing detector, see Fig.~\ref{fig:t_mean_model_and_measurement}.

These effects most likely also constitute the reason for the observed discrepancy, observed for $R^* > 50 \, \mathrm{MHz}$, between calibrated and inferred a priori detection rates, see Fig.~\ref{fig:inferred_rates}, as well as the deviation of fitted values of $\tauR$ in Fig.~\ref{fig:fit_stability}.

A potential explanation of these effects is presented below.

\begin{figure}[t]
    \includegraphics[width=\linewidth]{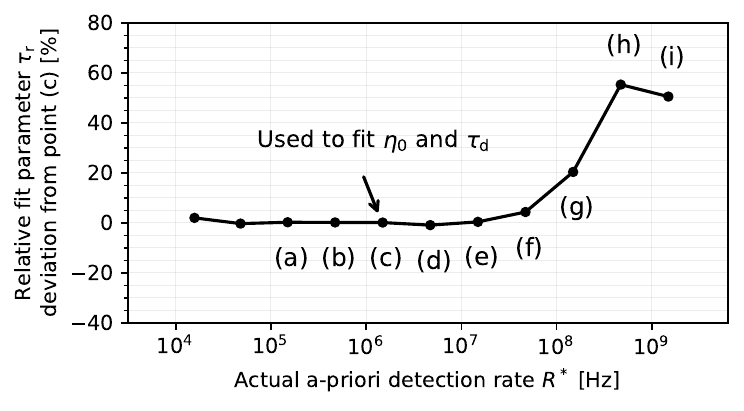}
    \caption{\label{fig:fit_stability}Fit stability.
    For each a priori detection rate $R^*$ the recovery-time constant $\tauR$ and a vertical scaling factor were fitted to the data.
    The parameters $\eta_0$ and $\tauD$ were fixed to the values obtained for point (c), see Fig.~\ref{fig:experiment_combined}(c).
    For a priori detection rates $R^* \le 49.2\,\mathrm{MHz}$ the fit parameters are very stable, confirming the assumption that $\tauR$ is an $R^*$-independent detector property.
    The labels refer to Fig.~\ref{fig:experiment_combined}.
    }
\end{figure}

\subsection{Paralyzing extension of the ER model}
\label{sec:paralyzing_model}

For $R^* \gtrsim 500 \, \mathrm{MHz}$ we observed falling detection rates $R$, suggesting that the SPAD exhibits a paralyzing behavior.
Therefore, a paralyzing extension of the ER model is introduced in the following.

We assume that avalanches triggered by a photon during a short paralyzable-time interval $\tauPOne$, right after the dead-time window, when the excess bias voltage is still very low, exhibit too small peak voltages to be registered by the latching circuit.
Hence, such avalanches do not result in a detection event and do not trigger a new dead-time window of duration $\tauD$.
However, the passive quenching circuit still leads to a lowering of the excess bias below threshold, prolonging the dead-time by another time constant $\tauPTwo$.
Furthermore, this effect can occur multiple times in a row.

\begin{figure}[t]
    \includegraphics[width=\linewidth]{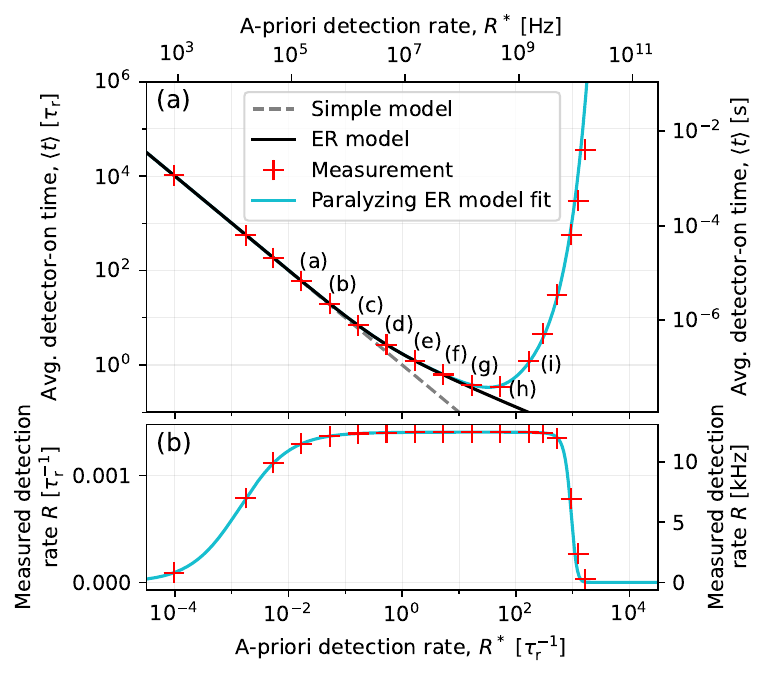}
    \caption{\label{fig:t_mean_model_and_measurement}Comparison of exponential-recovery (ER) model and experiment for the average detector-on time.
    The measurement points were calculated as $\langle t \rangle = 1/R - \tauD$, using the dead-time $\tauD = 80.092\,\mathrm{\mu s}$ as obtained from the fit shown in Fig.~\ref{fig:experiment_combined}(c).
    The ER model provides an accurate prediction of $\langle t \rangle$ for a priori detection rates up to $R^* \approx 10 \, \tauR^{-1}$.
    For higher values of $R^*$ additional effects emerge, cf.~Fig.~\ref{fig:experiment_combined}.
    To explain the observed reduction in detection rates $R$ for $R^* \gtrsim 10 , \tauR^{-1}$, we fitted a paralyzing extension of the ER model to the measured values of $\langle t \rangle$, see Section~\ref{sec:paralyzing_model}.
    To improve the fit around the minimum values of $\langle t \rangle$, data points were weighted by $1/\langle r \rangle^{3}$.
    }
\end{figure}

Using Eq.~\eqref{eq:cdf}, the probability for a paralyzation event to occur once is given by
\begin{align}
p_\mathrm{p}
= P(\tauPOne) \, .
\end{align}
Conditioned on such a paralyzation event happening, the average detector-on time is given by
\begin{align}
\langle t_\mathrm{p,1} \rangle = \frac{1}{P(\tauPOne)} \int_0^{\tauPOne} t \, p(t) \, \mathrm{d} t \, .
\end{align}
Hence, on average, every paralyzation event extends the dead-time by
\begin{align}
\langle t_\mathrm{p}^{(1)} \rangle
&= \langle t_\mathrm{p,1} \rangle + \tauPTwo \, .
\end{align}
Together with the average number of consecutive paralyzation events,
\begin{align}
\langle n \rangle
&= \sum_{k=0}^\infty k \, p_\mathrm{p}^k (1 - p_\mathrm{p})
= \frac{p_\mathrm{p}}{1-p_\mathrm{p}} \, ,
\end{align}
one finally finds the average total paralyzation duration
\begin{align}
\langle t_\mathrm{p}^\mathrm{(tot)} \rangle = \langle n \rangle \langle t_\mathrm{p}^{(1)} \rangle \, .
\end{align}
The paralyzing model average detector-on time is then given by
\begin{align}
\langle t_\textrm{paralyzing ER model} \rangle = \langle t_\textrm{ER model} \rangle  + \langle t_\mathrm{p}^\mathrm{(tot)} \rangle \, .
\end{align}
With the two fit parameters $\tauPOne$ and $\tauPTwo$, this model was fitted to the experimental data points, obtained as $\langle t \rangle = 1/R - \tauD$, using the previously fitted value for $\tauD$.
For the fit, the values for $\eta_0$, $\tauD$, and $\tauR$ were again fixed to the values obtained for Fig.~\ref{fig:experiment_combined}(c).

This fit resulted in $\tauPOne = 15.0 \pm 0.3 \, \mathrm{ns}$ and $\tauPTwo = 27 \pm 3 \, \mathrm{ns}$, and is in excellent agreement with the data, see Fig.~\ref{fig:t_mean_model_and_measurement}.
The value of $\tauPOne$ is roughly consistent with the duration of $10\,\mathrm{ns}$ after the dead-time window, during which we observed a deviation between model and data, described in Section~\ref{sec:limitations}, and visible in Fig.~\ref{fig:experiment_combined}.
Furthermore, the frequency $f \approx 40\,\mathrm{MHz}$ of the observed oscillations in the right tail of the PDF, see Fig.~\ref{fig:experiment_combined}(i), seems to be related to the corresponding value of the inverse of the average single-paralyzation prolongation of the dead-time, i.e., $\langle t_\mathrm{p}^{(1)} \rangle^{-1} \approx 28 \, \mathrm{MHz}$.
Combined with the overall good fit, shown in Fig.~\ref{fig:t_mean_model_and_measurement}, these two findings further support the paralyzing extension of the ER model.

%% file: 50_discussion.tex
\section{\label{sec:discussion}Conclusion}

The simple model, for instantaneous recovery after the dead-time, provides a good first order correction for the relation between a priori detection rates $R^* = R_\mathrm{i} \eta_0$ and measured detection rates $R$.
However, the simple model breaks down once $R^*$ becomes comparable to the inverse of the recovery time constant, $\tauR^{-1}$ (approx. $10 \, \mathrm{MHz}$ for our detector).

To improve the model accuracy, we derived an analytical model for SPADs with time-dependent exponential recovery of the quantum efficiency following dead-time.
This model is motivated by the behavior of typical electronic quenching circuits
\cite{covaAvalanchePhotodiodesQuenching1996},
and by previous observations of a time-dependent quantum efficiency
\cite{sanzaroInGaAsInPSPAD2016}.
A closed-form probability density function for the inter-detection interval was given in Eq.~\eqref{eq:exact_model_pdf}.
Our model only requires one additional parameter, the recovery time constant $\tauR$, which we established as an intrinsic detector property, independent of the impinging photon rate $R_\mathrm{i}$.
The ER model agrees with experimental data extremely well for a wide range of impinging photon rates, suggesting, that the exponential recovery is a good description for the underlying physical process.

Apart from the theoretical interest, our model has practical applications.
Note, that a single fit of the PDF $p(t)$ from Eq.~\eqref{eq:exact_model_pdf} to a measured inter-detection interval histogram suffices to obtain the asymptotic quantum efficiency $\eta_0$, dead-time $\tauD$, and exponential-recovery time constant $\tauR$, without requiring more complicated setups as, e.g., those employing a pulsed laser or an externally gated detector.
This simple method might be particularly useful for the characterization of SPADs
\cite{lopezStudyDevelopRobust2020}.

When using a SPAD to measure continuous-wave light power
\cite{lopezStudyDevelopRobust2020}, the simple model can lead to a significant underestimation of the impinging photon rate $R^*$, as shown in Fig.~\ref{fig:inferred_rates}.
In contrast, our ER model allows for a more precise calculation of the impinging photon rate, extending the applicable range in terms of $R^*$ by a factor of approx. 100.

For quantum key distribution (QKD)
\cite{pirandolaAdvancesQuantumCryptography2020},
the time-dependent quantum efficiency has implications on the security.
In case of slightly unequal dead-times or recovery time constants of the detectors used in a system, the resulting time-dependent detector efficiency mismatch leads to an advantage for an attacker Eve
\cite{
    makarovEffectsDetectorEfficiency2006,
    makarovPreparingCommercialQuantum2024,
    huangRealisticDetectorModel2025,
    georgievaDetectionUltraweakLaser2021}.
Furthermore, the paralyzation process described in Section~\ref{sec:paralyzing_model} could be exploited to selectively make individual detectors insensitive to impinging photons, which also provides an advantage to Eve
\cite{ImplementationAttacksQKD2023}.
It is most likely the underlying physical process of some of the reported detector blinding attacks against QKD
\cite{wuHackingSinglephotonAvalanche2020}.

To also capture the behavior for very high values of $R_\mathrm{i}$, we introduced a paralyzing model extension, which agrees with the data deep into the paralyzation regime.
It should be noted that the paralyzing model extension is non-invertible, such that the detection rate $R$ alone is not sufficient to unambiguously calculate the impinging photon rate $R_\mathrm{i}$.
However, this problem is general and inherent to all paralyzing detectors.

Our framework is generic and may equally assist disciplines that employ other dead-time-limited detectors, e.g., nuclear physics, particle physics, quadrupole mass spectrometry, or LIDAR 
\cite{
    gattDeadtimeEffectsGeigermode2007,
    liInfluenceDeadtimeDetection2017}.

Based on our general derivation from first principles, our model can also be adapted for other recovery functions $\eta(t)$.
Furthermore, by expressing the event rate as $\lambda(t) = \eta(t) R_\mathrm{i}(t)$, our framework can be extended for time-dependent light sources, as long as the time-dependent rate $R_\mathrm{i}(t)$ has a fixed temporal relation to the dead-time windows.
This might be of particular interest for TCSPC applications, where detection rates are limited by pile-up effects
\cite{liuFastFluorescenceLifetime2019}.
While this issue has been addressed in a recent publication
\cite{danieleBreakingBoundariesHybrid2025},
a combination with our model, and the non-homogeneous Poisson process framework in general, could potentially further improve the measurement precision for high detection rates.
Specifically, fit parameters of a general event rate function $\lambda(t)$ could be obtained, by fitting the PDF from Eq.~\eqref{eq:pdf_non_homogeneous_poisson} to the measured inter-detection interval histogram.

Future work could also explore more sophisticated paralyzing models.
While our paralyzing model extension fits the measured average inter-detection intervals very well, see Fig.~\ref{fig:t_mean_model_and_measurement}, the used fit parameters can only approximately explain the frequency of the observed oscillations in Fig.~\ref{fig:experiment_combined}(i).
These discrepancies could be addressed by incorporating factors beyond the scope of our model.
For example, the charge carrier extraction delay and the shape of the electrical avalanche pulse both depend on the time-dependent excess bias voltage.
Both effects should lead to a delayed detection for low excess bias voltages, distorting the inter-detection histograms for small $t$.
Further unconsidered effects include an overload of the SPAD readout electronics
\cite{saugeControllingActivelyquenchedSingle2011},
power supply overload
\cite{saugeControllingActivelyquenchedSingle2011},
and thermal blinding
\cite{lydersenThermalBlindingGated2010}.

For future work it is also of interest to incorporate afterpulsing
\cite{
    covaTrappingPhenomenaAvalanche1991,
    antiModelingAfterpulsingSinglephoton2011}.
In a first step, this could also be done via the framework of history-less non-homogeneous Poisson processes, used in this paper.
However, a more precise model, considering also the process history, should further improve the model accuracy, especially for short dead-times, where afterpulsing is more prominent.
Such models could be derived within the general framework of self-exciting point processes
\cite{
    snyderRandomPointProcesses1991,
    daleyIntroductionTheoryPoint2003,
    laubElementsHawkesProcesses2021},
e.g., via the Hawkes process, and could allow for even more precise detector characterizations, without requiring more complicated setups, like the double-gate method
\cite{stuckiPhotonCountingQuantum2001}.

In summary, our ER model provides a powerful analytical tool that not only improves the precision of photon flux estimation in dead-time-limited free-running SPADs, but also highlights the relevance of the non-homogeneous Poisson process for SPAD modeling, opening the door for further model refinements and a deepened understanding of detector physics.
Furthermore, due to its general formulation, our model is readily applicable to a broad class of other dead-time-limited detectors and a wide range of scientific disciplines.

%% file: 60_before_appendix.tex
\section*{\label{sec:supplementary_material}Supplementary Material}

See supplementary material [link to be added by journal] for a Python implementation of the probability density function of the ER model, which can readily be used for fitting experimental data.
Also, a numerical evaluation of the average detector-on time $\langle t \rangle$ is provided.

\section*{\label{sec:acknowledgements}Acknowledgements}

We thank Peter Hellwig for insightful discussions regarding the quenching electronics.
We thank Elisa Collin and Pascal Rustige for feedback on an early version of this manuscript.
ChatGPT was used during writing to polish and lightly edit some passages.
This research was conducted within the scope of the project QuNET+BlueCert, funded by the German Federal Ministry of Research, Technology and Space (BMFTR) in the context of the federal government's research framework in IT-security “Digital. Secure. Sovereign.”.

\section{Author Declarations}

\subsection*{Conflict of Interest}

The authors have no conflicts to disclose.

\subsection*{Author Contributions}

N.W. supervised the research and acquired the funding.
J.K. performed the research.
J.K. derived the theoretical models, conducted the experiments, evaluated the data, and wrote the paper.
All authors participated in discussions and reviewed the paper.

\textbf{Jan Krause:} Conceptualization (lead); Data curation (lead); Formal analysis (lead); Investigation (lead); Methodology (lead); Software (lead); Validation (lead); Visualization (lead); Writing - original draft (lead); Writing – review \& editing (lead).
\textbf{Nino Walenta:} Funding acquisition (lead); Project administration (lead); Supervision (lead); Validation (supporting); Writing – review \& editing (supporting)

\section*{Data Availability}

The data that support the findings of this study are available from the corresponding author upon reasonable request.

%% file: 70_appendix.tex
\appendix

\section{\texorpdfstring{ER model approximation for $R^* \ll \tauR^{-1}$}{ER model approximation for R* << 1/tau\_r}}
\label{app:exact_model_approximation_low_rates}

For small a priori detection rates, $R^* \ll \tauR^{-1}$, the PDF can be approximated by
\begin{align}
p(t)
\approx \big[ (1 + R^* \tauR) (1 - e^{-t/\tauR}) \big] \times R^* e^{-R^* t} \, ,
\end{align}
where a first order series expansion of the first factor of the correction term in Eq.~\eqref{eq:exact_model_pdf} was used.
This leads to an approximation of the expectation value
\begin{align}
\langle t \rangle
= \int_0^\infty t \, p(t) \, \mathrm dt
\approx \frac{1}{R^*} + \frac{\tauR }{1 + \tauR  R^*} \, .
\end{align}
Together with Eq.~\eqref{eq:general_rate_equation} this leads to the rate equations
\begin{align}
R
&\approx \frac{1}{1/R^* + \tauD + \left[ \frac{\tauR}{1 + \tauR R^*} \right] } \, ,
\end{align}
and
\begin{align}
R^* &\approx \frac{1}{1/R - \tauD} + \left[ \frac{1}{2 \tauR} \left( \sqrt{1 + \left( \frac{2 \tauR}{1/R - \tauD} \right)^2} - 1 \right) \right] \, ,
\end{align}
where again the square brackets are the corrections accounting for the exponential recovery, compared to Eqs.~\eqref{eq:simple_relation_1} and \eqref{eq:simple_relation_2}.

\section{\texorpdfstring{ER model approximation for $R^* \gg \tauR^{-1}$}{ER model approximation for R* >> 1/tau\_r}}
\label{app:exact_model_approximation_high_rates}

For large a priori detection rates, $R^* \gg \tauR^{-1}$, most detections occur for $t \ll \tauR$.
Therefore, the PDF from Eq.~\eqref{eq:exact_model_pdf} can be approximated by
\begin{align}
p(t)
\label{eq:approx_high_all}
&\approx \lambda^{(2)}(t) \exp \pmb{\Big(}- \int_0^t \lambda^{(2)}(\bar{t}) \, \mathrm{d}\bar{t} \pmb{\Big)} \, ,
\end{align}
where
\begin{align}
\lambda^{(2)}(t) = R^* \left[ \frac{t}{\tauR} - \frac{1}{2} \left( \frac{t}{\tauR} \right)^2 \right]
\end{align}
is the second order Taylor expansion of $\lambda(t) = R^* \eta(t)$ in $t$.
With another first-order expansion of the cubic term obtained after integration in Eq.~\eqref{eq:approx_high_all}, this leads to
\begin{align}
p(t)
&\approx R^* \left[ \frac{t}{\tauR} - \frac{1}{2} \left( \frac{t}{\tauR} \right)^2 \right] \exp \left( \frac{-R^* t^2}{2 \, \tauR} \right) \left( 1 + \frac{R^* t^3}{6 \, \tauR^2} \right) \, .
\end{align}
The average detector-on time is then given as
\begin{align}
\label{eq:approx_high_2}
\langle t \rangle
=
\sqrt{\frac{\pi \tauR}{2 R^*}} + \frac{1}{3 R^*} + \mathcal{O} \left( \frac{1}{R^* \sqrt{\tauR R^*}} \right) \, .
\end{align}
Taking only the first term of Eq.~\eqref{eq:approx_high_2} into account, gives the rate equations
\begin{align}
R &\approx
\frac{1}{\sqrt{\pi \tauR / 2 R^*} + \tauD} \, ,
\end{align}
and
\begin{align}
R^* &\approx
\frac{\pi \tauR / 2}{(1/R - \tauD)^2} \, .
\end{align}